# Viable thermionic emission from graphene-covered metals

E. Starodub, N. C. Bartelt and K. F. McCarty

Sandia National Laboratories, Livermore, California 94550, USA

Thermionic emission from monolayer graphene grown on representative transition metals, Ir and Ru, is characterized by low-energy electron microscopy (LEEM). Work functions were determined from the temperature dependence of the emission current and from the electron energy spectrum of emitted electrons. The high-temperature work function of the strongly interacting system graphene/Ru(0001) is sufficiently low, 3.3 ± 0.1 eV, to have technological potential for large-area emitters that are spatially uniform, efficient, and chemically inert. The thermionic work functions of the less strongly interacting system graphene/Ir(111) are over 1 eV larger and vary substantially (0.4 eV) between graphene orientations rotated by 30°.

The thermionic emission of electrons from a material is a phenomenon widely used in electronic technologies, particularly for electron emission.[1] Schottky-effect emitters rely on thermionic emission enhanced by an electric field.[2,3] Relatively efficient energy conversion using thermionic emission has been demonstrated [4] and an approach using photon enhancement has been proposed.[5] Key technological issues are developing emitters that are stable, insensitive to environment and are bright at low temperature, which minimizes the spread in the electron energy. Regarding brightness, thermionic emission becomes exponentially greater with decreasing work function, i.e., the energy needed to move an electron from the emitter material into the vacuum.

A standard approach to modifying a material's work function is to use molecular adsorbates to create surface dipoles.[6] Electropositive adsorbates lower the work function. The utility of this approach in practice is limited by the chemical reactivity and thermal stability of the adsorbates that significantly lower work function, such as cesium.[7] Graphene, which can be considered as a two-dimensional adsorbate, offers several advantages over molecular adsorbates for increasing thermionic emission. It is chemically inert, imparting resistance to molecular adsorption and oxidation.[8-10] Thus, robust emission is possible without ultrahigh vacuum environments. Graphene itself is stable to high temperatures. A single layer of graphene can be grown uniformly over very large areas of metal substrates,[11-14] giving spatially uniform, high-current emission. Alternatively, small, bright sources or arrays of sources can be patterned simply by removing the graphene using lithographic approaches. Graphene is known to lower the work function of metals at low temperature.[15-19] But at the elevated temperatures of thermionic emission, thermal excitations might change the weak binding of the film to the substrate, altering the work function and, consequently, the emission.

Here we investigate the viability of thermionic emission from graphene-covered metals by studying two representative transition metals, Ir and Ru. We measure the work function in two ways, from the temperature dependence of the emission current and from the energy spectrum of the emitted electrons. Graphene significantly lowers the work functions at high temperatures, as at low temperatures. The thermionic emission depends strongly on the strength of the graphene-metal interaction. In fact, the strongly interacting system graphene/Ru(0001) has a work function that is sufficiently low (3.3 eV) to have technological potential. The thermionic work functions of the less strongly interacting system graphene/Ir(111) are over 1 eV larger and vary substantially with the particular in-plane orientation of the graphene.

We studied the thermionic emission of graphene on metals using a low-energy electron microscopy (LEEM) instrument. Above about 1100K the current of thermionically emitted electrons is sufficiently large to image the emission from the surface. The power of this approach is that the emission intensity of precisely characterized regions (e.g., bare substrate vs. regions covered exactly by one graphene layer) can be measured from the images.[20] In fact, thermionic emission microscopy emerged before LEEM because of its greater simplicity (no independent electron source is needed).[21] The samples were held at -15 kV and separated by ~3 mm from a grounded objective lens, an electric field insufficient to generate Schottky emission. Temperature was measured from a thermocouple welded to the side of the Ru and Ir crystals, which were cleaned using procedures published elsewhere.[22,23] Graphene was grown on these metals by exposure to ethylene and by segregation from the bulk.[23,24] The quality of the graphene films was analyzed with low-energy electron diffraction (LEED). Thermionic electron energy distributions (TEED) of graphene-covered Ru(0001) were measured using a hemispherical energy analyzer integrated in the imaging system.[25]

The four images in Fig. 1 show thermionic emission as graphene grows on Ru(0001) by segregation during cooling. At the highest temperature (1240 K) the bright regions are covered by a single graphene layer while the dark regions are bare Ru. After cooling to 1200K the entire surface is covered by a single layer. The image intensity decreases with decreasing temperature. Cooling below 1160K leads to the nucleation of the second graphene layer, which images darker because of its higher work function. The graphene film has high quality, as demonstrated by the sharpness of the high-order superstructure spots in the LEED pattern of Fig. 1 (b). Fig. 1 also plots the thermionic emission as a function of temperature. The dependence is strong, with emission decreasing by about a factor of six with a temperature drop of 250K. We fit our experimental data to the Richardson-Dushman equation[26] for the emission current density $I=AT^2 exp(-\phi/kT)$, where $A$ is a material-specific constant, $T$ is the sample temperature, $\phi$ is the work function, and $k$ is Boltzmann's constant. The equation is derived by assuming that emitted electrons originate from an ideal electron gas whose energies are described by the Maxwell-Boltzmann distribution and that the work function $\phi$ is independent of temperature. For graphene/Ru(0001) we find $\phi$ = 3.3 ± 0.1 eV, where the error is the standard deviation of five measurements. This value is considerably smaller than the work functions of either Ru(0001) or graphite, 5.3 eV [27] and 4.83 eV,[28] respectively. The measured value agrees well with the result of first-principles calculations by Wang et al., 3.6 eV.[29] These calculations show that the relatively strong graphene/Ru interaction leads to significant film-to-substrate charge transfer, which creates a surface dipole that lowers the work function.[17] The reasonable agreement between the 0K calculation and the high-temperature measurement shows that the charge transfer is not sensitive to temperature.

The extent of charge transfer between graphene and the metal, i.e., the strength of the surface dipole, is determined by the graphene/metal interaction. For example, graphene binds more weakly to the Ir(111) surface[30,31] than to Ru(0001).[12,29] Fig. 2 compares the thermionic emission from Ru(0001) (red circles) and Ir(111) (blue squares) both covered by monolayer graphene that has the same azimuthal crystallographic orientation. We call this in-plane orientation, in which the graphene and metal lattices are aligned, R0.[32] Fits to the Richardson-Dushman equation gives $\phi_{R0/Ir(111)}$ = 4.2 ± 0.1eV, where the error is the standard deviation of four measurements. Thus, covering Ir reduces its work function (5.8 eV)[33] by 1.6 eV, which is less than the 2 eV reduction caused by covering Ru. This result is consistent with calculations[29,30] and photoemission experiments[19,34] that show stronger hybridization in the graphene/Ru system.

While graphene only occurs as a single orientation on Ru(0001), multiple in-plane orientations can occur on metals that interact more weakly, such as Ir(111).[19] We show next that these orientations can drastically affect thermionic emission. Fig. 3 (a) shows an image formed by electrons emitted at 1450K from three types of Ir(111) regions – bare substrate, covered by R0 graphene and covered by graphene that is rotated 30°, R30. (The LEED patterns in Fig. 3 (b) and (c) document assignment of the orientations.) The black regions are the bare Ir, which emits essentially no electrons at this temperature. Dramatic contrast exists between the two different graphene orientations − the R0 regions are bright and the R30 regions are gray. Thus, thermionic emission is very sensitive to in-plane azimuth. Fig. 3 (a) also highlights how uniform the emission is within a rotational domain. On average there are only small modulations in intensity, which result from the topography of the substrate. There are, however, a low density of brighter emission points (small white regions) that result from bunches of substrate steps. An example is the circular emission site at the bottom, which likely results from a closed loop of bunched substrate steps.

Fig. 2 compares the temperature dependence of the emission from the R0 (blue squares) and R30 orientations (green triangles) on Ir(111). The separation of the curves quantifies the large effect of in-plane orientation. Fitting to the Richardson-Dushman equation gives $\phi_{R30/Ir(111)}$ = 4.6 ± 0.1 eV, 0.4 eV higher than $\phi_{R0/Ir(111)}$. Our previous investigation by angle-resolved photoemission of the effect of in-plane orientation on electronic structure[19] revealed that the less-stable R30 variant had weaker hybridization with Ir bands near the Fermi level than did R0. The larger work function of R30 graphene is consistent with this orientation having smaller charge transfer to the metal.[19]

We further explored thermionic emission from graphene/Ru(0001) using an independent technique – from the energy distribution of the thermionic electrons. The LEEM's spectrometer was calibrated using secondary electrons photo-excited from a W(110) single crystal by a helium lamp. The cutoff in the secondary electron distribution was taken to be the W work function measured by thermionic emission (5.25 eV)[26,35-38]. The thermionic electron energy distribution (TEED) of graphene-covered Ru(0001) at 1300K is shown with red circles in Fig. 4 (a). The full width at half maximum (FWHM) of the peak, ~0.38 eV, is comparable with the width (0.33 eV) of the LEEM's $LaB_6$ emitter, whose work function we measure to be 2.9 eV (see Fig. 4 (b)), close to the literature value of 2.8 eV.[15] To correct for instrumental broadening, we fit the experimental data to a convolution of Maxwell-Boltzmann and Gaussian distributions. Four parameters were fit: the energy shift of the Maxwell-Boltzmann distribution, the sample temperature $T$, the standard deviation of the Gaussian distribution $\sigma = \text{FWHM}/\sqrt{8ln2}$, and a multiplication constant. The fit, the solid black line in Fig. 4 (a), gives $\phi_{G/IRu(0001)} = 3.300 \pm 0.002$ eV, where the error is the standard deviation of the fit, $T$=1450K and $\sigma$=0.081 eV. The goodness of the fit establishes that the thermionic electrons indeed have a Maxwell-Boltzmann energy distribution. We attribute the discrepancy between the fitted and measured temperature to inaccuracies in the temperature measurement. Note that the work functions measured by TEED (Fig. 4) and from the temperature dependence (Figs. 1 and 2) agree.

In summary, the temperature dependence and the energy distribution of thermionic emission from Ir and Ru covered by graphene monolayers are described well by the classical description. The work function of Ru(0001) covered by a single graphene layer is remarkably low at high temperature, 3.3 ± 0.1 eV. In contrast, the less strongly bound system, graphene-covered/Ir(111), has a larger work function, 4.4 eV on average. Moreover, graphene's in-plane orientation on Ir(111) strongly affects work function, causing a difference of 0.4 eV. The strong binding between graphene and Ru(0001) yields a low work function and a single in-plane azimuthal orientation, which in turn gives uniform emission. Together with graphene's inertness, these properties give graphene/Ru(0001) technological potential as a large-area emitter.

This work was supported by the Office of Basic Energy Sciences, Division of Materials Sciences and Engineering of the US DOE under Contract No. DE-AC04-94AL85000.

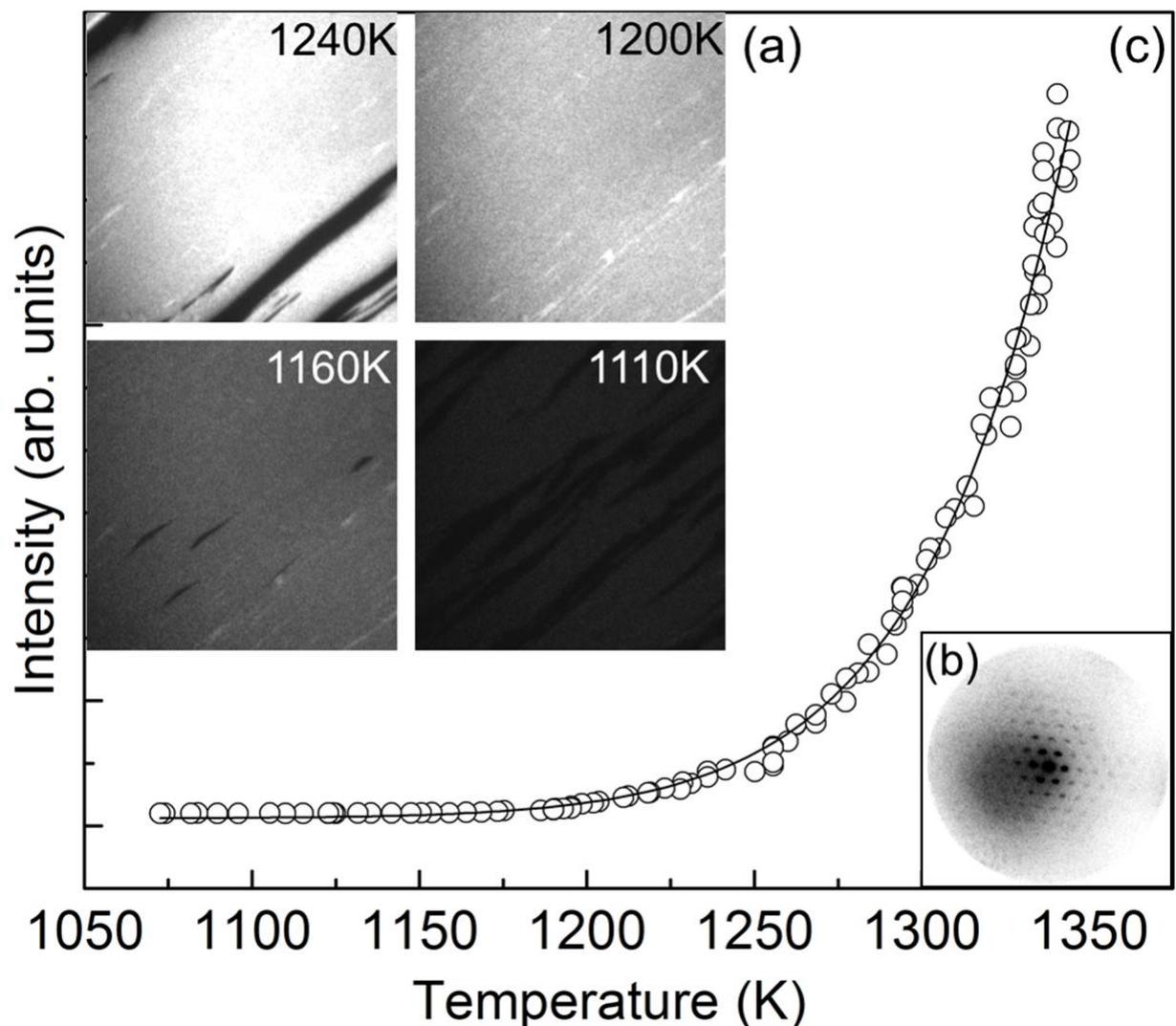

Fig.1 (a) Thermionic emission images of graphene-covered Ru(0001) at indicated temperatures. The field-of view is 25μm × 25μm. At 1240K, the dark and bright contrasts are bare Ru and graphene, respectively. At 1160 and 1110K, the dark islands are two-layer graphene. (b) LEED pattern of the graphene-covered Ru(0001) showing the superstructure spots around the specular beam at 33 V. (c) Intensity of the thermionically emitted electrons from the graphene-covered Ru(0001) as a function of temperature (black circles) and a fit to the Richardson-Dushman equation (solid line).

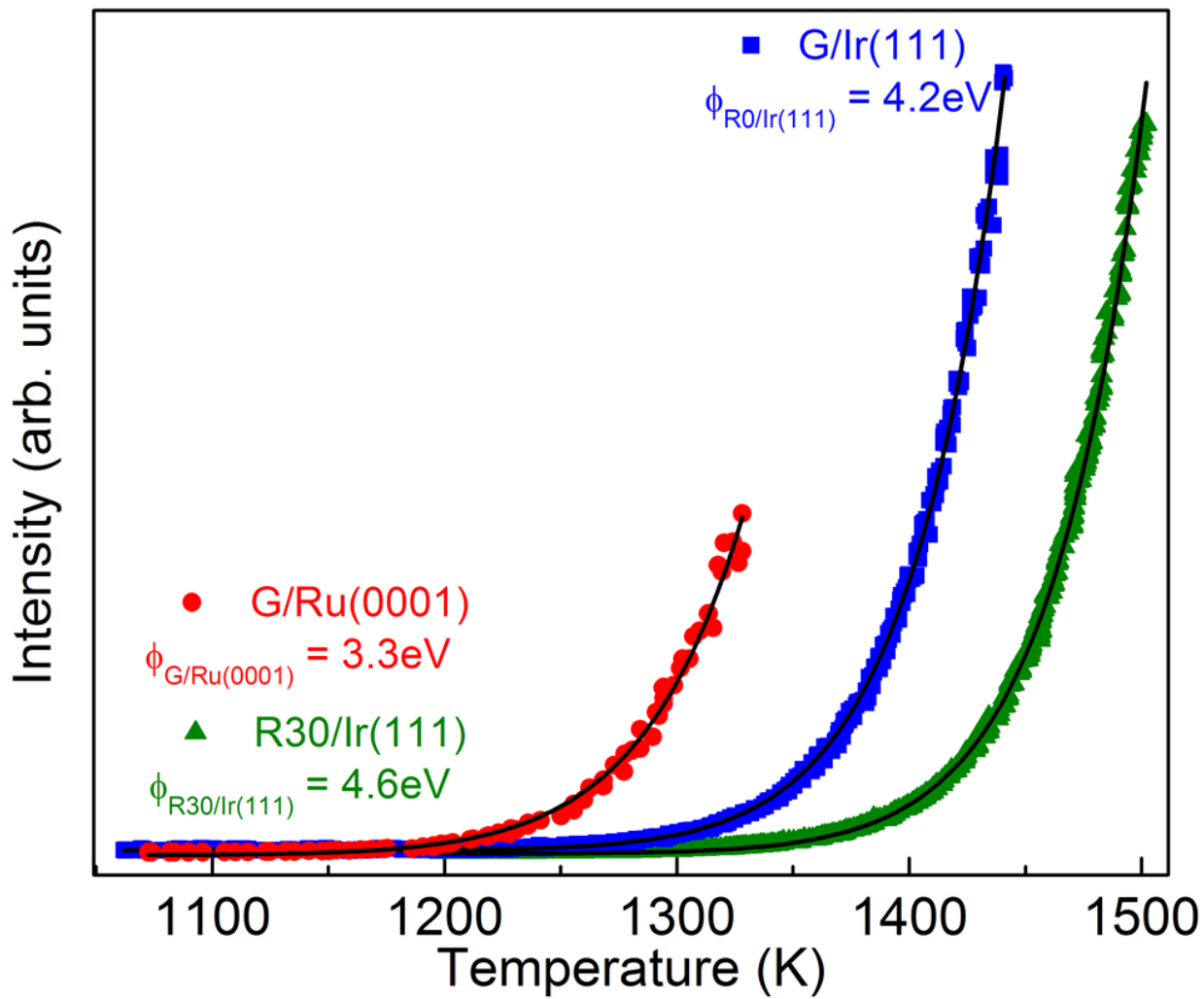

Fig.2 (Color online) Temperature-dependence thermionic emission from Ru(0001) (red circles) and Ir(111) covered by a single graphene layer. Two in-planes orientations are measured on Ir(111, R0 (blue squares) and R30 (green triangles). Solid lines are fits to the Richardson-Dushman equation.

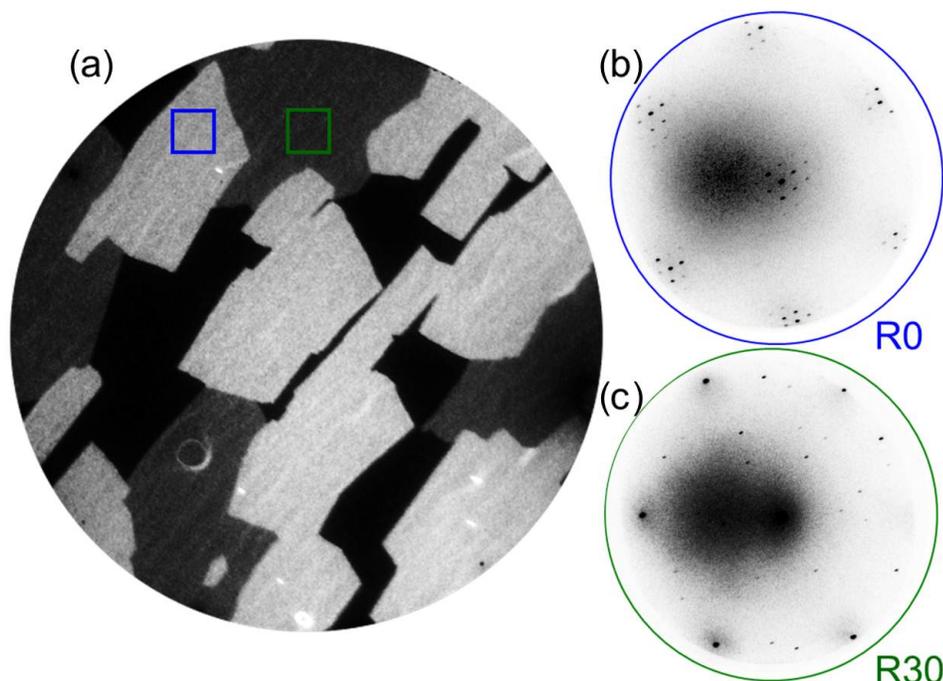

Fig.3 (Color online) (a) Thermionic emission image of graphene islands on Ir(111) at 1450K. The field-of-view is 93μm × 93μm. The bright and gray contrasts are single-layer islands with R0 and R30 orientation, respectively. The black regions are bare Ir. (b) and (c) LEED patterns (40 eV) from the regions marked with the blue and green boxes, respectively.

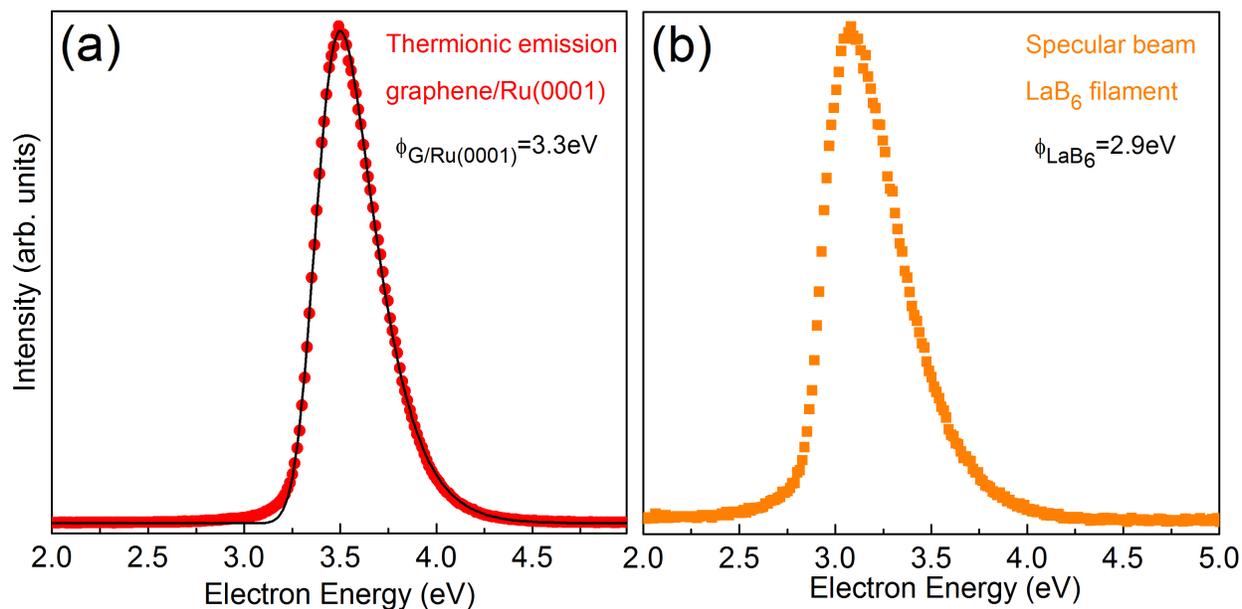

Fig.4 (Color online) (a) Thermionic electrons energy distribution (TEED) (red circles) from Ru(0001) covered by a single graphene layer at 1300K. The solid line is the fit to a convolution of Maxwell-Boltzmann and Gaussian distributions. (b) Energy distribution from the $LaB_6$ emitter of the LEEM.